\documentclass[11pt]{article}
\usepackage{epsfig}
\usepackage{a4p}

\begin{document}

\newcommand {\Zzero}     {\rm{Z}}
\newcommand {\MZ}        {m_{\mathrm{Z}}}
\newcommand {\MH}        {m_{\mathrm{H}}}
\newcommand {\MT}        {m_{\mathrm{t}}}
\newcommand {\MW}        {m_{\mathrm{W}}}
\newcommand {\GZ}        {\Gamma_{\mathrm{Z}}}
\newcommand {\Gl}        {\Gamma_{\ell}}
\newcommand {\Ge}        {\Gamma_{\mathrm{e}}}
\newcommand {\Gb}        {\Gamma_{\mathrm{b}}}
\newcommand {\Gc}        {\Gamma_{\mathrm{c}}}
\newcommand {\Gf}        {\Gamma_{\mathrm{f}}}
\newcommand {\Ghad}      {\Gamma_{\rm{had}}}
\newcommand {\sighad}    {\sigma_{\mathrm{had}}}
\newcommand {\sigzhad}   {\sigma^0_{\mathrm{had}}}
\newcommand {\sigzf}     {\sigma^0_{\mathrm{f}}}
\newcommand {\sigee}     {\sigma_{\mathrm{ee}}}
\newcommand {\sigmu}     {\sigma_{\mu\mu}}
\newcommand {\sigtau}    {\sigma_{\tau\tau}}
\newcommand {\sigl}      {\sigma_{\ell^+\ell^-}}
\newcommand {\Rl}        {R_{\ell}}
\newcommand {\Rb}        {R_{\mathrm{b}}}
\newcommand {\Afb}       {A_{\mathrm{FB}}}
\newcommand {\Afbee}     {A^{\mathrm{e}}_{\mathrm{FB}}}
\newcommand {\Afbmu}     {A^{\mu}_{\mathrm{FB}}}
\newcommand {\Afbtau}    {A^{\tau}_{\mathrm{FB}}}
\newcommand {\Afbzl}     {A^{0,\ell}_{\mathrm{FB}}}
\newcommand {\Afbzb}     {A^{0,\mathrm{b}}_{\mathrm{FB}}}
\newcommand {\Afbzc}     {A^{0,\mathrm{c}}_{\mathrm{FB}}}
\newcommand {\Afbzf}     {A^{0,\mathrm{f}}_{\mathrm{FB}}}
\newcommand {\Ae}        {{\cal A}_{\mathrm{e}}}
\newcommand {\Atau}      {{\cal A}_{\tau}}
\newcommand {\Al}        {{\cal A}_{\ell}}
\newcommand {\Ab}        {{\cal A}_{\mathrm{b}}}
\newcommand {\Ac}        {{\cal A}_{\mathrm{c}}}
\newcommand {\Af}        {{\cal A}_{\mathrm{f}}}
\newcommand {\Rc}        {R_{\mathrm{c}}}
\newcommand {\Wpm}       {\mathrm{W}^{\pm}}
\newcommand {\WpWm}      {\mathrm{W}^+\rm{W}^-}
\newcommand {\WW}        {\mathrm{WW}}
\newcommand {\GW}        {\Gamma_{\mathrm{W}}}
\newcommand {\ee}        {\mathrm{e}^+\mathrm{e}^-}
\newcommand {\mumu}      {\mu^+\mu^-}
\newcommand {\tautau}    {\tau^+\tau^-}
\newcommand {\elel}      {\ell^+\ell^-}
\newcommand {\ff}        {\mathrm{f\bar{f}}}
\newcommand {\qq}        {\mathrm{q\bar{q}}}
\newcommand {\bb}        {\mathrm{b\bar{b}}}
\newcommand {\cc}        {\mathrm{c\bar{c}}}
\newcommand {\pp}        {\mathrm{p\bar{p}}}
\newcommand {\gamgam}    {\gamma\gamma}
\newcommand {\Ebm}       {E_{\mathrm{beam}}}
\newcommand {\alfamz}    {\alpha(\MZ)}
\newcommand {\alfasmz}   {\alpha_s(\MZ)}
\newcommand {\GF}        {G_F}

\newcommand {\thw}        {\theta_{\mathrm{W}}}
\newcommand {\thweff}     {\theta_{\mathrm{W}}^{eff}}
\newcommand {\theffl}     {\theta_{\mathrm{eff}}^{\mathrm{lept}}}
\newcommand {\swsq}       {\sin^2\!\thw}
\newcommand {\swsqweff}   {\sin^2\!\thweff}
\newcommand {\swsqeffl}   {\sin^2\!\theffl}
\newcommand {\cwsq}       {\cos^2\!\thw}

\newcommand {\gvl}        {g_{V\ell}}
\newcommand {\gal}        {g_{A\ell}}
\newcommand {\gve}        {g_{V\mathrm{e}}}
\newcommand {\gae}        {g_{A\mathrm{e}}}
\newcommand {\gvf}        {g_{V\mathrm{f}}}
\newcommand {\gaf}        {g_{A\mathrm{f}}}
\newcommand {\gle}        {g_{L\mathrm{e}}}
\newcommand {\gre}        {g_{R\mathrm{e}}}
\newcommand {\glf}        {g_{L\mathrm{f}}}
\newcommand {\grf}        {g_{R\mathrm{f}}}

\newcommand {\sigf}       {\sigma_{\rm{F}}}
\newcommand {\sigb}       {\sigma_{\rm{B}}}

\newcommand {\sigleft}    {\sigma_{L}}
\newcommand {\sigright}   {\sigma_{R}}

\newcommand {\Alr}        {A_{LR}}
\newcommand {\Afblr}      {A_{FBLR}}

\newcommand {\thetavec} {\mbox{\boldmath$\theta$\unboldmath}}
\newcommand {\thmunu}    {\theta_{\mu\nu}}
\newcommand {\ce}        {c_{\eta}}
\newcommand {\se}        {s_{\eta}}
\newcommand {\cx}        {c_{\xi}}
\newcommand {\sx}        {s_{\xi}}

\newcommand {\cz}        {c_{\zeta}}
\newcommand {\sz}        {s_{\zeta}}
\newcommand {\ca}        {c_{\delta}}
\newcommand {\sa}        {s_{\delta}}
\newcommand {\cd}        {c_{\alpha}}
\newcommand {\sd}        {s_{\alpha}}
\renewcommand {\d}         {\mbox{d}}

\newcommand {\LNC}       {\Lambda_{\rm NC}}

\newcommand {\dravsn}     {Final DRAFT}
\newcommand {\dradate}    {\today}
\newcommand {\commdate}   {5 March, 2003}
\newcommand {\commtime}   {12:00}
\newcommand {\commentto}  {Tatsuo.Kawamoto@cern.ch and Kirsten.Sachs@cern.ch}


\flushbottom
\begin{titlepage}
%
%
\begin{center}{\Large
EUROPEAN ORGANISATION FOR NUCLEAR RESEARCH
}\end{center}\bigskip
\begin{flushright}
       CERN-EP/2003-010 \\ 5 March 2003
\end{flushright}
%
%
\bigskip
\boldmath
\begin{center}{\huge\bf
Test of non-commutative QED \\
in the process $\ee\rightarrow\gamgam$ at LEP \\
}
\end{center}\unboldmath\bigskip
\vspace*{1cm}
\begin{center}{\LARGE The OPAL Collaboration
}\end{center}\bigskip\bigskip
%
%
\bigskip\bigskip
%
%


%
\vspace*{0.3cm}
%
%
\begin{center}{\large Abstract}\end{center}

Non-commutative QED would lead to deviations from the Standard Model 
depending on a new energy scale $\LNC$ and a unique direction in
space defined by two angles $\eta$ and $\xi$. Here in this analysis
$\eta$ is defined as the angle
between the unique direction and the rotation axis of the earth.
The predictions of such a theory for the process 
$\ee\rightarrow\gamma\gamma$ are evaluated for the specific orientation 
of the OPAL detector and compared to the measurements.
Distributions of the polar and azimuthal scattering angles are
used to extract limits on the energy scale $\LNC$ 
depending on the model parameter $\eta$. 
At the 95\% confidence level $\LNC$ is found to be larger than 141 GeV
for all $\eta$ and $\xi$. It is shown that the
time dependence of the total cross-section could be used
to determine the model parameter $\xi$ if there were a 
detectable signal. These are the first limits obtained on 
non-commutative QED from an $\rm e^+e^-$ collider experiment.

%
%
%
%
%
\bigskip\bigskip\bigskip\bigskip
\bigskip\bigskip
\begin{center}{\large
To be submitted to Phys. Lett.}\end{center}
%
%

\end{titlepage}

\begin{center}{\Large        The OPAL Collaboration
}\end{center}\bigskip
\begin{center}{
G.\thinspace Abbiendi$^{  2}$,
C.\thinspace Ainsley$^{  5}$,
P.F.\thinspace {\AA}kesson$^{  3}$,
G.\thinspace Alexander$^{ 22}$,
J.\thinspace Allison$^{ 16}$,
P.\thinspace Amaral$^{  9}$, 
G.\thinspace Anagnostou$^{  1}$,
K.J.\thinspace Anderson$^{  9}$,
S.\thinspace Arcelli$^{  2}$,
S.\thinspace Asai$^{ 23}$
% ...
}\end{center}\bigskip
$^{  1}$School of Physics and Astronomy, University of Birmingham,
Birmingham B15 2TT, UK
\newline
$^{  2}$Dipartimento di Fisica dell' Universit\`a di Bologna and INFN,
I-40126 Bologna, Italy
\newline
$^{  3}$Physikalisches Institut, Universit\"at Bonn,
D-53115 Bonn, Germany
\newline
% ...
$^{  5}$Cavendish Laboratory, Cambridge CB3 0HE, UK
\newline
% ...
$^{  9}$Enrico Fermi Institute and Department of Physics,
University of Chicago, Chicago IL 60637, USA
\newline
% ...
$^{ 16}$Department of Physics, Schuster Laboratory, The University,
Manchester M13 9PL, UK
\newline
% ...
$^{ 22}$Department of Physics and Astronomy, Tel Aviv University,
Tel Aviv 69978, Israel
\newline
$^{ 23}$International Centre for Elementary Particle Physics and
Department of Physics, University of Tokyo, Tokyo 113-0033, and
Kobe University, Kobe 657-8501, Japan
\newline
% ...
$^{  b}$ and Royal Society University Research Fellow
\newline
$^{  c}$ and Institute of Nuclear Research, Debrecen, Hungary
\newline
$^{  d}$ and Heisenberg Fellow
\newline
$^{  e}$ and Department of Experimental Physics, Lajos Kossuth University,
 Debrecen, Hungary
\newline
$^{  f}$ and MPI M\"unchen
\newline
% ...
$^{  p}$ now at Physics Dept Southern Methodist University, Dallas, TX 75275,
U.S.A.
\newline
$^{  q}$ now at IPHE Universit\'e de Lausanne, CH-1015 Lausanne, Switzerland
\newline
% ...
$^{  t}$ now at RWTH Aachen, Germany

\clearpage

\section{Introduction} 

Recently, there has been increasing interest in theories with non-commutative 
space-time geometries. The idea of non-commutative geometry is not new.
It was studied in the 1940s as a possible
means of regularising divergences in quantum field theory~\cite{snyder}.
More recent interest is related to the possibility that non-commutative 
geometry may arise in string theory through the quantisation of strings 
in the presence of background fields~\cite{b-string}.

In a quantum field theory of non-commutative geometry the space-time 
coordinates are represented by operators $X_{\mu}$ satisfying 
the relation:
\begin{equation}
\label{eq-commu}
[X_{\mu}, X_{\nu}]=i\thmunu
\end{equation}
where $\thmunu$ is a constant antisymmetric matrix, having units of
(length)$^2$ = (mass)$^{-2}$ $\sim$ $1/\LNC^2$. This introduces a 
fundamental scale, $\LNC$, representing the space-time distance
below which the space-time coordinates become fuzzy.
Its role can be compared to that of the Planck constant $h$ in 
ordinary quantum mechanics, which quantifies the level of non-commutativity
between coordinates and momenta.l
Although there is no {\it a priori} prediction for the scale $\LNC$, 
it might 
be at the level of the Planck scale 
\cite{b-ncqedrev,b-ncqedscale}.
However, in light of recent progress in string theory and theories 
with large extra dimensions the energy scale at which gravity
becomes strong could be of ${\cal O}(\mbox{TeV})$. It is therefore
conceivable that $\LNC$ 
could also be at the TeV scale and that the effects of a non-commutative geometry 
might be observable in present or planned collider experiments.

The matrix $\thmunu$ can be decomposed into two independent parts
\cite{b-ncqedph2,b-kamoshita}:
electric-like components 
$\thetavec_E = (\theta_{01}, \theta_{02}, \theta_{03})$
and magnetic-like components 
$\thetavec_B = (\theta_{23}, \theta_{31}, \theta_{12})$.
The matrix $\thmunu$ is constant and frame independent, which leads to 
violation of Lorentz invariance; $\thetavec_E$ and 
$\thetavec_B$ can be 
considered as 3-vectors which define two unique directions in space.

A non-commutative Standard Model has not yet been formulated.
Only QED with non-commu\-ta\-tive geometry (NCQED) exists~\cite{b-ncqed}.
This theory is known to be invariant under U(1) gauge transformations 
and renormalisable at the one-loop level. However, there are some limitations in 
the existing NCQED, for example only charges 0 or $\pm 1$ are allowed
and consequently quarks are not incorporated in the theory. 
Despite these limitations NCQED can be regarded 
as a consistent theory, and may serve as a test bed for the study of 
other non-commutative quantum field theories.
There are a number of studies of general NCQED 
phenomenology~\cite{b-ncqedrev} and specifically at high energy 
linear colliders~\cite{b-ncqedph2, b-ncqedph3}.
Limits have been obtained at low energy using 
the Lamb shift \cite{b-ncqed-lamb}, 
the Aharonov-Bohm effect \cite{b-ncqed-AB} and
clock comparisons \cite{b-ncqed-clock} under specific assumptions.
This paper presents the first limits on NCQED obtained from a collider
experiment.

In NCQED each ee$\gamma$ vertex involves a kinematic phase factor 
$e^{\frac{i}{2}p_1^{\mu}\thmunu p_2^{\nu}}$, where 
$p_1^{\mu}$, $p_2^{\nu}$ are the electron momenta.
In addition there are non-Abelian-like 3$\gamma$ and 4$\gamma$ 
self couplings, which are proportional to the kinematic phase.
The amplitude of a scattering process therefore depends not only on the
kinematics of the initial and final state particles, but also on the 
unique directions $\thetavec_E$ and $\thetavec_B$ 
relative to the orientation of the experiment. 
The characteristic properties of NCQED may thus be observed as 
direction-dependent deviations
from the predictions of QED. If observed, the unique directions could 
be inferred. This can be regarded as an analogue of the 
Michelson-Morley experiment.

In this paper, a purely electromagnetic process
$\ee\rightarrow\gamgam$ is studied using the high-energy 
$\ee$ collision data collected with the OPAL detector at LEP. 
The theoretical calculation corresponds to tree level 
($\ee\rightarrow\gamgam$). The experimental selection includes 
higher orders ($\ee\rightarrow\gamgam(\gamma)$)
and the measured cross-sections are corrected to tree level assuming
ordinary QED. Possible higher-order effects from NCQED are expected
to be smaller than effects of fourth order QED and weak interactions
which are taken into account by a 1\% systematic uncertainty on the
cross-section.

\section{\boldmath $\ee\rightarrow\gamgam$ in NCQED \unboldmath}

In NCQED three diagrams contribute to the process 
$\ee\rightarrow\gamma\gamma$ at the tree level.  Two are similar to the 
ordinary pair-annihilation diagrams of QED, but with a kinematic phase at 
each vertex. The third diagram is an $s$-channel photon exchange with 
$\gamma\gamma\gamma$ self coupling.
The differential cross-section for $\ee\rightarrow\gamgam$ in NCQED is 
given by~\cite{b-kamoshita}
\begin{equation}
\label{eq-dsiggg}
\frac{\d^2\sigma}{\d\cos\theta \d\phi} = 
\frac{\alpha^2}{s}
\frac{1+\cos^2\theta}{1-\cos^2\theta}
[1-\sin^2\theta\cdot\sin^2\Delta_{\rm NC}], 
\end{equation}
where $\theta$ and $\phi$ are the polar and azimuthal angles
(with respect to the outgoing electron beam) of the final state 
photon with $0\leq\cos\theta\leq 1$,
$\alpha$ is the fine-structure constant and 
$s$ is the centre-of-mass energy squared.
This is similar to the QED expression but with a correction
arising from NCQED represented by the term in square brackets.
The parameter $\Delta_{\rm NC}$ is given by
\begin{equation}
\label{eq-delnc}
\Delta_{\rm NC}=-\;\frac{s}{4\LNC^2}
(c_{01}\sin\theta\cos\phi + c_{02}\sin\theta\sin\phi + c_{03}\cos\theta).
\end{equation}
Here we have introduced new dimensionless parameters, $c_{0i}$, 
defined by 
\begin{equation}
\label{eq-cedef}
\thetavec_E = \frac{1}{\LNC^2}{\bf c }_E = 
\frac{1}{\LNC^2}(c_{01}, c_{02}, c_{03})
\end{equation}
where $\LNC=1/\sqrt{|\thetavec_E|}$ and 
$c_{0i}$ are components of the unit vector ${\bf c}_E$
pointing to the unique
direction in the coordinate system of the experiment.
At tree level the process $\ee\rightarrow\gamgam$ is sensitive 
only to $\thetavec_E$, not to $\thetavec_B$.
For final state photons which are not back-to-back small effects from
the magnetic-like components $\thetavec_B$ could occur which are
neglected in this analysis.
Note also that the effect of NCQED on the cross-section for this process is 
always negative with respect to QED and that the relative size of the effect 
is larger at large production angles due to the $\sin^2\theta$ term.
In general $\Delta_{\rm NC}$ depends not just on the photon production angle
$\theta$, but also on the azimuthal angle $\phi$.
This is a signature of the anisotropy of space-time which is inherent in 
non-commutative geometry. Only in the special case where $\thetavec_E$ 
is parallel to the beam electrons
($c_{01} = c_{02} = 0$) does this $\phi$ dependence vanish.
Even in the presence of transverse beam polarisation the QED 
cross-section is independent of $\phi$ \cite{b-page}.

The unique direction ${\bf c}_E$ is not known.
However, if it exists, it is unlikely that it is fixed to the solar system 
or to the earth.
Rather it would be natural to assume that this direction is fixed 
to some larger structure in space, e.g.\ the rest frame of 
the cosmic microwave background. We refer to this as the primary frame.
In the coordinate system of an experiment on the earth, 
the unique direction will change as the earth rotates and as the orientation
of the earth's rotation axis changes due to the movement of the galaxy or 
the solar system with respect to the primary frame. We assume that the 
latter movement is sufficiently slow that over the timescale
of the experiment the rotation of the earth is the only 
relevant motion. This in turn provides an opportunity to examine 
the time-dependent effect of NCQED.
In the next Section, we consider how the direction $c_{0i}$ varies
as a function of the earth's rotation and how $\Delta_{\rm NC}$ follows its 
variation.

\section{\boldmath Vector $\thetavec_E$ in the experimental 
laboratory system \unboldmath}

\begin{figure}[tbp]
\mbox{\epsfxsize1.00\textwidth\epsffile{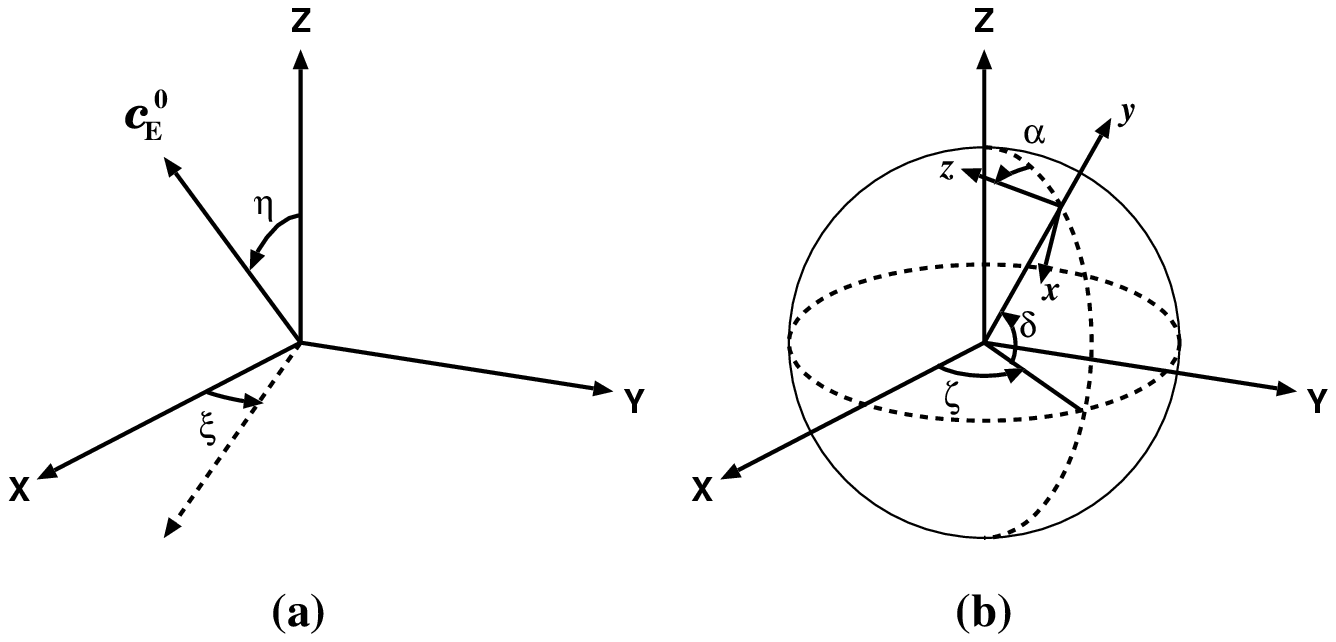}}
\caption{\label{f-axes} Definition of the two coordinate systems:
(a) the primary frame ($X, Y, Z$) in which the vector ${\bf c}^0_E$ is fixed, and
(b) the local coordinate system ($x, y, z$) of an $\ee$ experiment on the earth.
}
%
\centerline{\mbox{\epsfxsize0.8\textwidth\epsffile{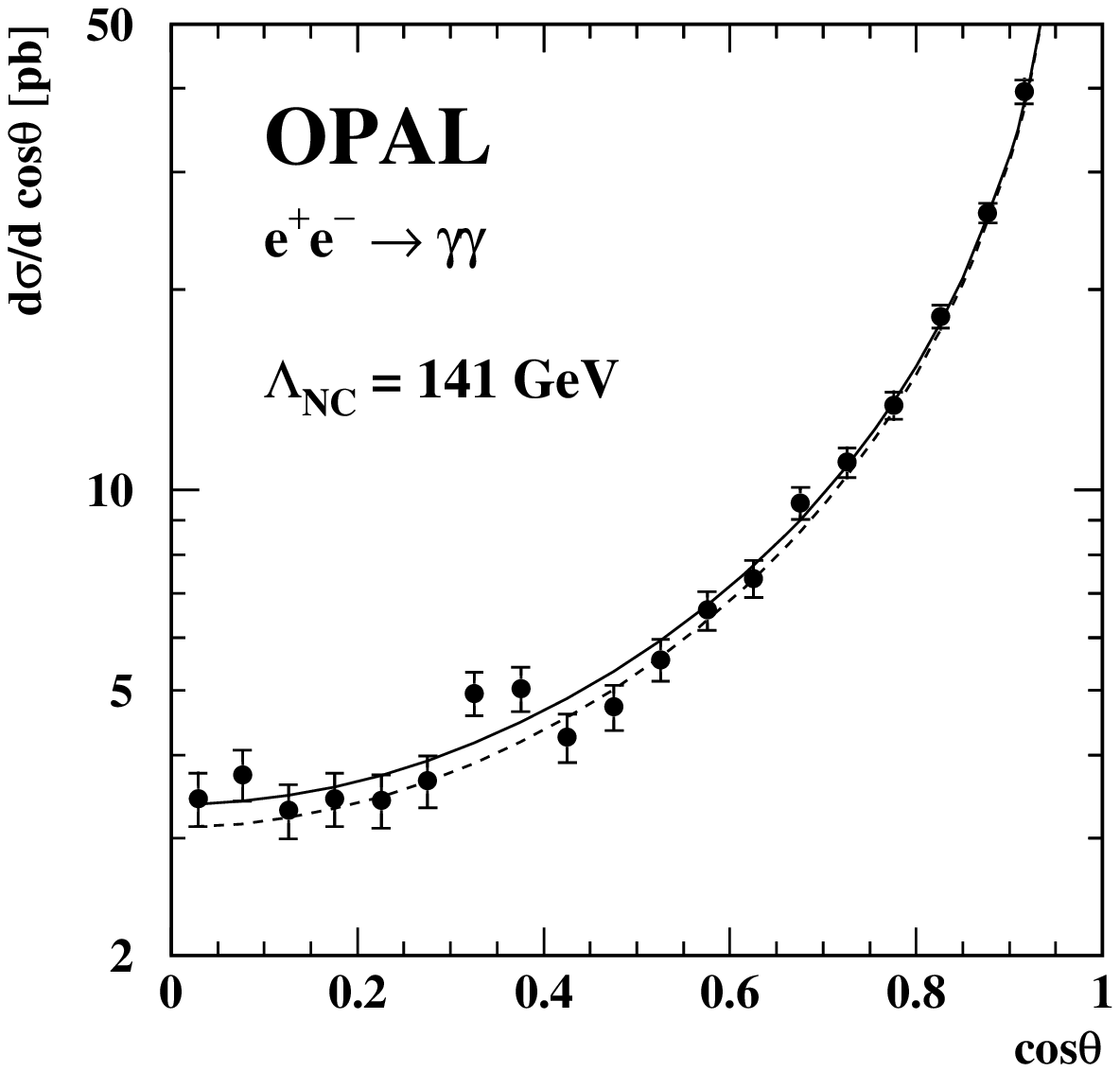}}}
\caption{\label{fig:ct} 
The measured differential cross-section as a function of 
$\cos{\theta}$. The points are OPAL data, the solid line shows 
the Standard Model prediction and the dashed line corresponds to 
the 95\% confidence level limit of $\LNC = 141 \mbox{ GeV}$
and $\eta = 90^\circ$.
}
\end{figure}

The conventions for the primary $(X,Y,Z)$ and local $(x,y,z)$ coordinate systems  
are shown in Figure~\ref{f-axes}. We use right-handed Cartesian coordinate systems 
throughout this paper.
For convenience, we choose the axis of the earth's rotation to be the 
$Z$-axis of the primary coordinate system. The
$X$-axis points in some fixed direction which can be chosen arbitrarily.
The unit vector ${\bf c}^0_E$ in the primary frame is specified 
by two parameters, the polar angle $\eta$ and the azimuthal angle $\xi$: 
\begin{equation}
{\bf c}^0_E = 
    \left(\begin{array}{r}
      \se\cx \\ \se\sx \\ \ce 
      \end{array}\right),
\end{equation}
where $\se = \sin\eta$, $\cx = \cos\xi$ and so on.
 
On the earth, the experiment is located at a point of latitude $\delta$.
The local coordinate system $(x,y,z)$ is defined such that the
$z$-axis is in the direction of the e$^-$ beam which is in a horizontal plane
with respect to the surface of the earth, 
the $y$-axis is vertical and the $x$-axis is perpendicular to the $y$-$z$ plane.
This is, to a good 
approximation, the same definition as used in the OPAL experiment\footnote{
The LEP ring is not precisely horizontal but it is tilted by 0.8$^{\circ}$.
However, for the purposes of this analysis this angle is small and 
will thus be neglected.
}.
The angle between the $z$-axis and the direction of north is denoted by 
$\alpha$ (measured counter-clockwise, see Figure~\ref{f-axes}).
As the earth rotates, the local coordinate system moves around the $Z$-axis.
The time-dependent azimuthal angle, $\zeta(t)$, of the location of the experiment 
with respect to the $X$-axis is given by,
\begin{equation}
\zeta = \omega t, 
\end{equation}
where $\omega=2\pi/T_{sd}$ with sidereal day $T_{sd}$,
the time taken for one complete rotation of the earth around its axis.

Elements of the vector ${\bf c}_E$ in the local coordinate system $(x, y, z)$
are obtained by successive rotations of the coordinate axes~\cite{b-kamoshita}:
\begin{equation}
{\bf c}_E = R\cdot {\bf c}^0_E
\end{equation}
\begin{equation}
R = R_y(\alpha) R_z(-\pi/2) R_y(-\delta) R_z(\zeta)
\end{equation}
yielding
\begin{eqnarray}
\label{eq-ce}
{\bf c}_E & = & 
    \left(\begin{array}{r}
      \sd\sa \\ \ca \\ -\cd\sa 
      \end{array}\right) \se\cdot\cos(\zeta - \xi) 
~~ + ~~\left(\begin{array}{r}
       \cd \\ 0 \\ \sd 
      \end{array}\right) \se\cdot\sin(\zeta - \xi)  
~~ + ~~\left(\begin{array}{r}
      -\sd\ca \\ \sa \\ \cd\ca  
      \end{array}\right) \ce.  
\end{eqnarray}
The elements in the brackets form three constant vectors which are determined 
by the location of the accelerator and orientation of the e$^-$ beam at
the experiment. The
coefficients of these three constant vectors depend on the unique direction
in the primary frame ($\eta$ and $\xi$) and the phase of the earth's rotation
$\zeta$.
There are two distinct components, one which varies with time and one which is constant.
When the angle $\eta =0$, i.e.\ the vector ${\bf c}_E$ is parallel 
to the rotation axis of the earth, the time-dependent component vanishes.
The $\phi$ dependence of $\Delta_{\rm NC}$ (Equation~\ref{eq-delnc}) 
in general exists, but vanishes in the special configuration where 
$\eta=0$ and the experiment is located on the equator with e$^-$ beam pointing 
to the north ($\alpha=0$, $\delta=0$) or south.
The OPAL experiment is located at the latitude $\delta=$ 46.29$^{\circ}$ N 
and the longitude of 6.11$^{\circ}$ E with the angle $\alpha = 33.69^{\circ}$
\cite{b-position}.

In this analysis, we choose the $X$-axis to point in the
direction of the vernal equinox in the Pisces constellation.
Due to the precession of the earth, and other reasons, the direction of
the vernal equinox varies with time over a period of many years. However, 
for the limited duration of LEP operation, its motion can be neglected. 
In this approximation, the angle $\zeta$ at 
time~$t$ is given by 
\begin{equation}
\zeta = \frac{2\pi}{T_{sd}}(t-t_0) + \zeta_0
\end{equation}
where $T_{sd}$ is the average sidereal day (23h 56min 4.09053sec) \cite{b-sday}, 
and $\zeta_0$ is the azimuthal angle of the LEP location at time $t_0$,
chosen to be the moment of vernal equinox in 1995:  $t_0$ = 21st~March~1995, 
02h~14min~(UT) \cite{b-equinox}.  The angle $\zeta_0$ at that time is $\zeta_0 = 219.6^{\circ}$.
Each OPAL event has a time stamp, so the angle $\zeta$ can be
determined for each event. 

In summary, for the OPAL experiment, 
\begin{eqnarray}
\alpha= 33.69^{\circ}         \hspace{4mm} ; \hspace{4mm}
\delta= 46.29^{\circ}         \hspace{4mm} ; \hspace{4mm}
\zeta = \frac{2\pi}{T_{sd}}(t-t_0) + \zeta_0 
\end{eqnarray}
while the direction of the ${\bf c}_E$ vector ($\eta$ and $\xi$)
and the energy scale $\LNC$ are unknown.

\section{Data analysis}

To search for the effects of NCQED we use events selected as 
$\ee\rightarrow\gamma\gamma (\gamma)$ as described in detail 
in~\cite{b-multiphoton}. The events were collected from a data sample
corresponding to an integrated luminosity of 672.3~pb$^{-1}$ taken
at the highest LEP centre-of-mass energies, between 181 GeV and 209 GeV, 
during the last four years of  OPAL operation. The luminosity-weighted 
mean centre-of-mass energy is 196.6~GeV. 
In total, 5235 events with at least two photons observed in the region 
$|\cos{\theta_\gamma}| < 0.93$ are selected, where $\theta_\gamma$ is
the photon angle. No restrictions on either the angle between the 
two highest-energy photons or the number of additional photons 
are applied. The estimated background is less than 
0.3\%. The excellent uniformity and hermeticity of the OPAL detector 
provide a high and uniform efficiency of 98\% over the entire range of
azimuthal 
angle $\phi$ and for $|\cos{\theta_\gamma}| < 0.80$. Only in the range
$0.80 < |\cos{\theta_\gamma}| < 0.93$ do larger corrections have to be applied.
The experimental systematic uncertainties are small, typically 0.8\%.
The error on the theoretical prediction of the Standard Model QED
is assumed to be 1\%. 
This theoretical error arises from the correction of the observed
angular distribution to the Born level at which the model predictions 
and measured cross-sections are given. 
This correction involves the angular definition of events with a
topology other than two back-to-back photons. As in 
\cite{b-multiphoton} the event polar angle convention chosen is
$\cos{\theta} = \left|\sin{\frac{\theta_1 - \theta_2}{2}}\right|
 \;  / \; \left( {\sin{\frac{\theta_1 + \theta_2}{2}}}\right)$,
where $\theta_1$ and $\theta_2$ are the polar angles of the two
highest-energy photons. The event azimuthal angle $\phi$ is chosen to be
the azimuthal angle of the photon with largest $\cos{\theta_\gamma}$ 
out of the two highest-energy photons.

In NCQED the differential cross-section depends on three kinematic variables:
the polar angle $\theta$, the azi\-mu\-thal angle $\phi$, and the time
via the orientation angle $\zeta$. For each of these three variables, a fit 
to the model prediction is performed with the cross-section integrated 
(or averaged) over the other two. For example, the $\theta$ dependence is studied with a 
distribution integrated over $\phi$ and averaged over $\zeta$. Similarly, 
the total cross-section, integrated over $\theta$ and $\phi$, is analysed as
a function of $\zeta$. Because the region of large $\cos{\theta}$ is
dominated by the Standard Model, the relative effects of NCQED are enhanced by 
restricting the $\theta$ integration to $\cos{\theta}<0.6$, leading 
to a subsample of 1800 events. 

The luminosity delivered by LEP is not uniformly distributed over time. 
On average more data were collected at night than during the day,
leading to a dependence of the luminosity on $\zeta$ with a
variation of 13\%. This effect is taken into account in the
determination of the cross-sections. The total integrated
luminosity is determined using small-angle Bhabha scattering 
in the region $25 \mbox{ mrad} < \theta < 59 \mbox{ mrad}$.
The $\zeta$ dependence of the luminosity is obtained from the rate of 
Bhabha events detected in the electromagnetic calorimeter
($|\cos{\theta}|<0.96$) in the same data sample as used 
for the analysis. Here we assume that any NCQED effects on Bhabha scattering 
can be neglected since the cross-section is dominated by forward scattered 
events for which
the effects of NCQED are expected to be small~\cite{b-ncqedph2}.

To obtain a log likelihood curve which is approximately
parabolic a fit parameter $\varepsilon$ is chosen such that 
$\LNC = (|\varepsilon|)^{-1/4}$.
The non-physical region of negative $\varepsilon$ is included in the
fit by replacing $\sin^2{\Delta_{\rm NC}}$ by 
$- \; \sin^2{\Delta_{\rm NC}}$ in Equation \ref{eq-dsiggg}
if $\varepsilon$ is negative. Integration and averaging of this 
cross-section is done numerically.

\subsection{\boldmath The $\cos\theta$ distribution \unboldmath}

Although the NCQED contribution $\Delta_{\rm NC}$ depends on several 
unknown model parameters ($\LNC$, $\eta$, $\xi$),
the dependence on $\eta$ and $\xi$ is greatly reduced 
when the cross-section is integrated over $\phi$ and averaged over time 
(i.e.\ $\zeta-\xi$).
In this case, the only relevant kinematic variable in the differential 
cross-section is the production polar angle $\theta$, and deviations
from QED depend mainly on the unknown 
parameter $\LNC$. For the orientation of OPAL the 
dependence on the parameter $\eta$ is very weak.
The effects are largest at $\cos{\theta}=0$, here the variation
of the cross-section with $\eta$ is $\sim 0.2$\%. 
Therefore the result obtained on $\LNC$ from 
the $\cos\theta$ distribution is almost independent of $\eta$.
However, this decoupling is accidental; for other values of $\alpha$
the $\cos\theta$ distribution might vary by up to 40\% relative
to the average as a function of $\eta$.

The measured $\cos\theta$ distribution, at the luminosity-weighted 
average centre-of-mass energy, is shown in Figure \ref{fig:ct}.
This distribution is sensitive to the parameter $\LNC$.  
Since no significant deviation from QED is observed a limit on $\LNC$
is set, assuming $\eta=90^{\circ}$. This yields the most conservative 
result, since for this value of $\eta$ the difference between QED 
and NCQED for a given $\LNC$ is smallest.
To obtain this limit, a simultaneous binned log-likelihood fit is 
performed to the differential 
cross-section distributions measured at eight centre-of-mass energy points 
and including systematic uncertainties and their correlations in the manner 
described in \cite{b-multiphoton}. The result of this fit is given in 
Table~\ref{tab:fitres}. The corresponding one-sided 
limit at 95\% confidence level of $\LNC > 141~\mbox{GeV}$ 
shown in Figure \ref{fig:limit} as the dark grey region
is valid for all~$\eta$ and, obviously, for all $\xi$.

\subsection{\boldmath The $\phi$ distribution \unboldmath}

\begin{figure}[tbp]
\vspace*{-2cm}
\centerline{\mbox{\epsfxsize0.8\textwidth\epsffile{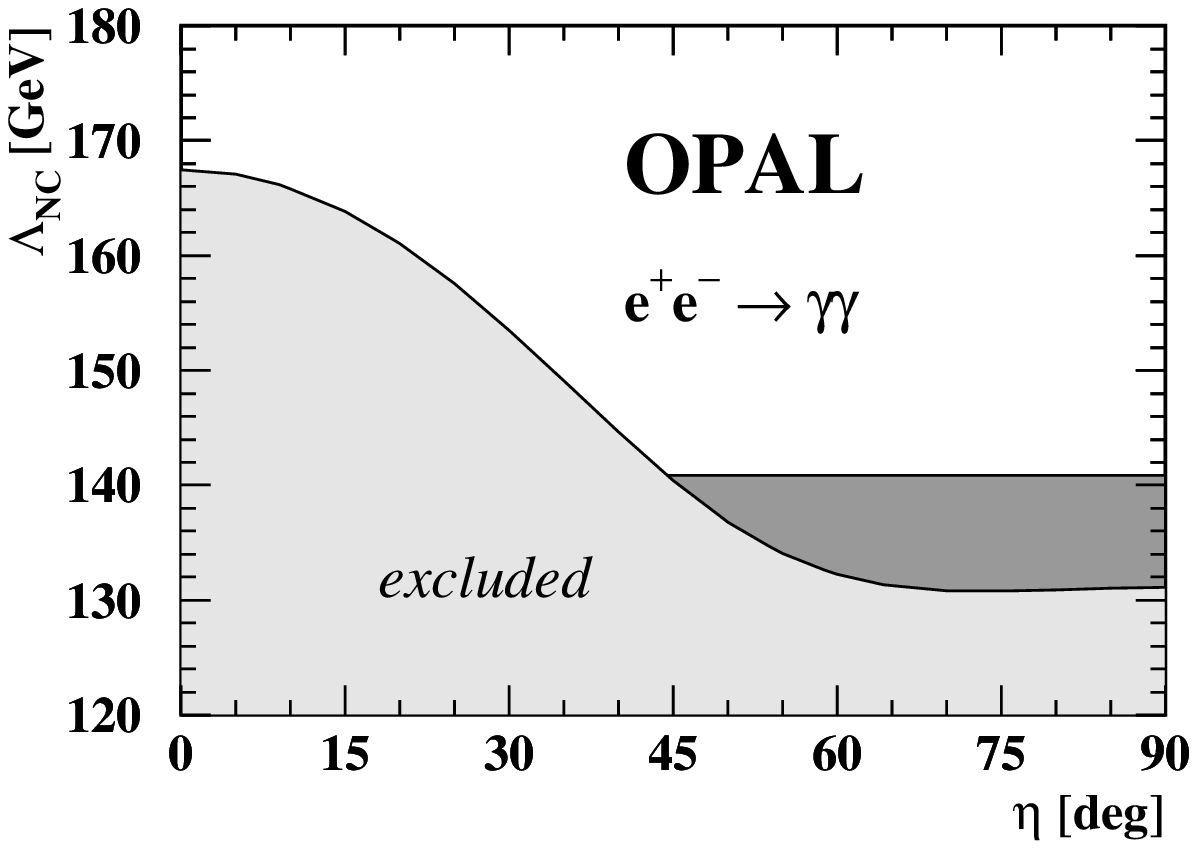}}}
\caption{\label{fig:limit} 
One-sided 95\% confidence level limits on the energy scale
$\LNC$ as a function of the angle $\eta$.
The light grey region is derived from the 
observed $\phi$ distribution and the $\eta$-independent limit shown 
in dark grey results from the observed $\cos{\theta}$ distribution. 
}

\centerline{\mbox{\epsfxsize0.8\textwidth\epsffile{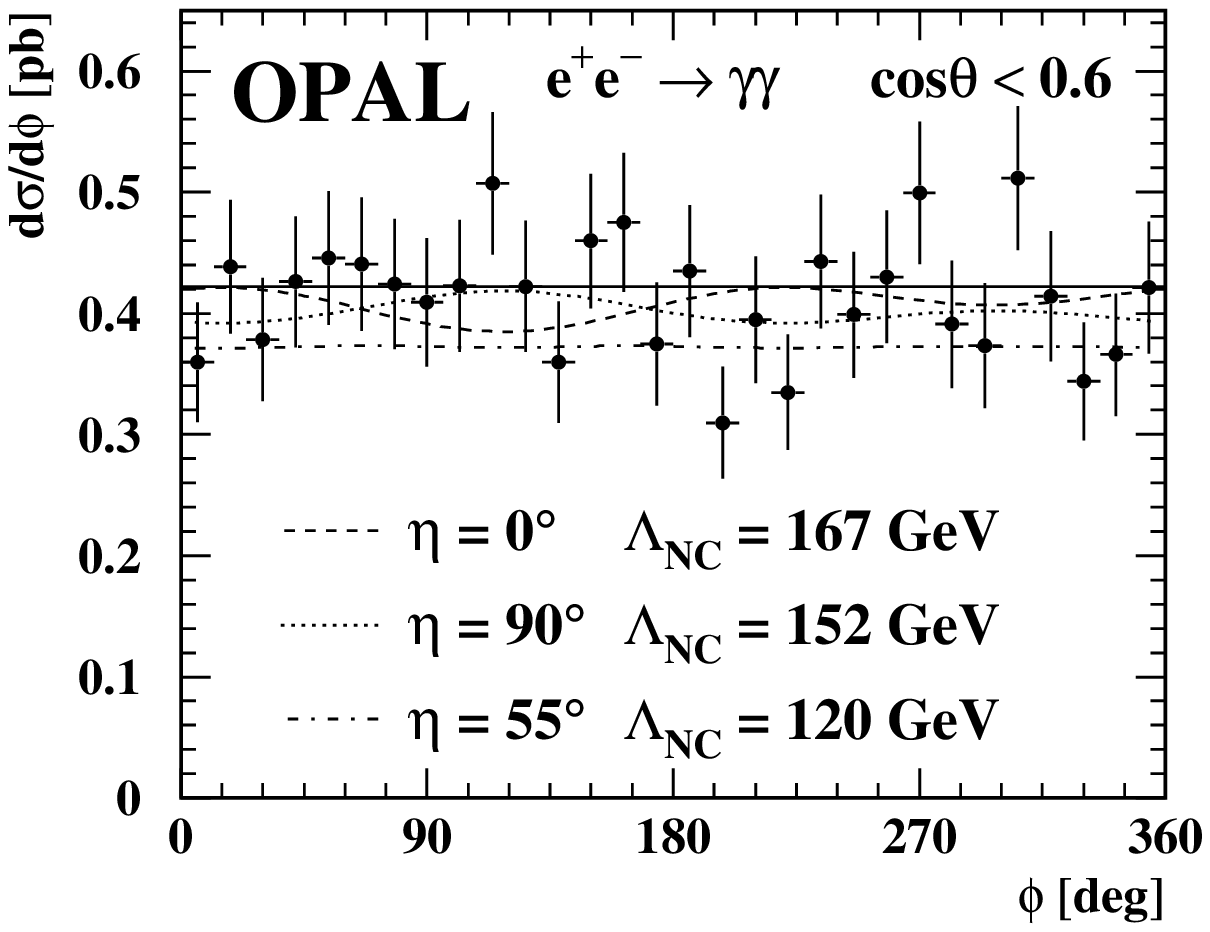}}}
\caption{\label{fig:phi} 
Measured $\phi$ distribution.
The points are OPAL data and the solid line the Standard
Model prediction. Expectations from NCQED are shown for the best limit
of $\LNC > 167 \mbox{ GeV}$ at $\eta = 0^\circ$ and the best fit
at $\eta = 90^\circ$ which yields $\LNC = 152 \mbox{ GeV}$.
The $\phi$ independent distribution at $\eta = 55^\circ$ is shown as
well.
}
\end{figure}

As discussed above, a $\phi$-dependent cross-section is a 
characteristic signature of NCQED.
The $\phi$-dependence arises from the $x$ and $y$ components 
$(c_{01}, c_{02})$ of the vector ${\bf c}_E$.
Since these components have a time-independent component 
(Equation~\ref{eq-ce}), some $\phi$-dependence remains even if the data 
are averaged over all $\zeta$. However, for certain values of $\eta$, 
depending on the orientation of the experiment, the $\phi$-dependence
is washed out completely.
In the configuration of the OPAL experiment at LEP, such a cancellation 
occurs for \mbox{$\eta\approx$ 55$^{\circ}$} (or 125$^{\circ}$) as can be 
seen in Figure \ref{fig:phi}.

\begin{table}[tp]
\begin{center}
\renewcommand{\arraystretch}{1.5}
\begin{tabular}{l|c|c}\hline
Fit  & Fit result [TeV$^{-4}$] & 95\%\ CL Limit [GeV]\\  \hline\hline
$\cos{\theta}$ distribution &  & \\[-1ex]
$\eta = 90^\circ$ & $947 {+920 \atop -905}$ & 141 \\ \hline
$\phi$ distribution & &  \\[-1ex]
$\eta = 0^\circ$ & $-174 {+703 \atop -732}$ & 167 \\ 
$\eta = 30^\circ$ & $225 {+836 \atop -841}$ & 154 \\ 
$\eta = 60^\circ$ & $1589 {+999 \atop -982}$ & 132 \\ 
$\eta = 90^\circ$ & $1866 {+913 \atop -900}$ & 131 \\ \hline\hline
\end{tabular}
\end{center}
\caption{\label{tab:fitres}
Results of the fits to the $\cos{\theta}$ distribution and the $\phi$
distribution at several values of $\eta$.
The fit result is given for the fit parameter $\varepsilon$.
For $\LNC$ the one-sided limit at 95\% confidence level is given.} 
\end{table}

Figure~\ref{fig:phi} shows the measured $\phi$ distribution, at the 
luminosity-weighted average centre-of-mass energy, for 
$\ee\rightarrow\gamma\gamma$ integrated over $\cos\theta<0.6$
and averaged over time. From this distribution 95\% confidence level
lower limits on $\LNC$ are obtained as a function of $\eta$ and
independent of $\xi$. 
Table~\ref{tab:fitres} summarises the fit results four four
values of $\eta$. For $\eta = 90^\circ$ the best fit is two standard
deviations away from the Standard Model. The result of that fit with a 
central value of $\LNC = 1866^{-1/4} \mbox{ TeV} = 152 \mbox{ GeV}$, 
as well as the limit for $\eta = 0^\circ$ are shown in 
Figure \ref{fig:phi}. For illustration the $\phi$ independent expectation
at $\eta = 55^\circ$ at a low scale of $\LNC=120 \mbox{ GeV}$ is added
to the plot.  Limits for all values of $\eta$ are shown in 
Figure~\ref{fig:limit} as the light grey region.
The limit is strongest for $\eta = 0^\circ$ and weakest for 
$\eta = 73^\circ$.  For some values of
$\eta$ the modulation in $\phi$ is small and sensitivity is lost
due to the integration over $\cos{\theta}$, leading to a large
uncertainty on $\varepsilon$. A better strategy would
be a two-dimensional fit to $\d^2\sigma / \d\cos{\theta} \d\phi$. This 
analysis could be performed in several bins of $\zeta$ to avoid the loss of
information due to the integration over time. However, such an analysis 
is not feasible given the available data statistics.

\subsection{The total cross-section}

The limits given above are obtaind from time-averaged distributions and
hence give no information about the third model parameter $\xi$.
This information could be provided by the total cross-section which
depends on $\zeta-\xi$. 
However, any limit on $\LNC$ determined from the total cross-section
in dependence of $\xi$ would be weaker than the limits given above,
since the differential cross-sections in $\cos{\theta}$ or $\phi$ provide 
the strongest sensitivity to the scale $\LNC$.
But if a signal were observed, the unique time structure of the
total cross-section would allow the determination of the angle $\xi$.

\begin{figure}[tbp]
\centerline{\mbox{\epsfxsize0.8\textwidth\epsffile{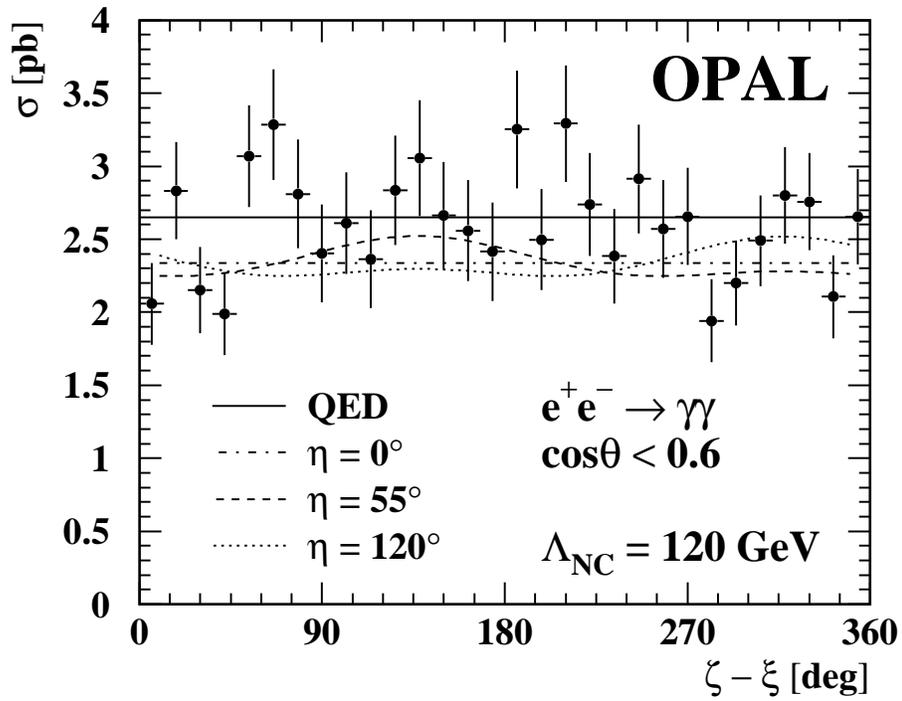}}}
\caption{\label{fig:zeit} 
Total cross-section as a function of the time-dependent $\zeta-\xi$.
The points are OPAL data and the solid line represents the Standard
Model prediction. To guide the interpretation of the data the 
expectations from NCQED
are shown for a low scale of $\LNC = 120 \mbox{ GeV}$. Three 
values of $\eta$  (0$^\circ$, 55$^\circ$, 120$^\circ$) are shown.
}
\end{figure}

Figure \ref{fig:zeit} shows the total cross-section, at the 
luminosity-weighted centre-of-mass energy, integrated over
$\theta$ and $\phi$, as a function of the time-dependent angle~$\zeta$.
The distribution has a $\chi^2 / \mbox{dof} = 39.9/30$ with respect to 
the Standard Model expectation which corresponds to a probability of 11\%. 
Since no signal is observed in either 
the $\cos{\theta}$ or the $\phi$ distribution, no attempt is made to extract 
$\xi$ from a fit to the measured total cross-section. However, examples
of two model expectations are shown in Figure \ref{fig:zeit} together
with the time independent case of $\eta = 0^\circ$.

\section{Conclusion}

Non-commutative QED would lead to deviations from the Standard Model
depending on some energy scale $\LNC$ and on a unique direction in 
space given by the angles $\eta$ and $\xi$. The experimental signature 
of such a theory for the process $\ee\rightarrow\gamma\gamma$ 
was evaluated for the orientation of the OPAL detector 
and compared to the measurements. No significant deviations from the 
Standard Model predictions were observed.
Distributions of the scattering angle $\theta$ were used 
to extract a lower limit on the energy scale $\LNC$ of 141 GeV at the 
95\% confidence level, which is valid for all angles $\eta$ and $\xi$.
Using the $\phi$ distributions this limit was improved for some 
values of $\eta$ up to $\LNC > 167$ GeV. The total cross-section 
gives weaker limits but its time dependence could in principle be used
to determine the third model parameter $\xi$ if a signal were observed. 
These are the first limits
obtained on NCQED from an $\rm e^+e^-$ collider experiment.

\section*{Acknowledgements}

We would like to thank J. Kamoshita and K. Hagiwara for 
useful discussions. \\
We particularly wish to thank the SL Division for the efficient operation
of the LEP accelerator at all energies
 and for their close cooperation with
our experimental group.  In addition to the support staff at our own
institutions we are pleased to acknowledge the  \\
Department of Energy, USA, \\
National Science Foundation, USA, \\
Particle Physics and Astronomy Research Council, UK, \\
Natural Sciences and Engineering Research Council, Canada, \\
Israel Science Foundation, administered by the Israel
Academy of Science and Humanities, \\
Benoziyo Center for High Energy Physics,\\
Japanese Ministry of Education, Culture, Sports, Science and
Technology (MEXT) and a grant under the MEXT International
Science Research Program,\\
Japanese Society for the Promotion of Science (JSPS),\\
German Israeli Bi-national Science Foundation (GIF), \\
Bundesministerium f\"ur Bildung und Forschung, Germany, \\
National Research Council of Canada, \\
Hungarian Foundation for Scientific Research, OTKA T-029328, 
and T-038240,\\
The NWO/NATO Fund for Scientific Reasearch, the Netherlands.\\


\clearpage

\end{document}